\documentclass{article} % For LaTeX2e

\usepackage{paper_style,times}

\ClearShipoutPicture

\usepackage{amsmath,amsfonts,bm}

\def\eqref#1{equation~\ref{#1}}
\def\1{\bm{1}}

\DeclareMathAlphabet{\mathsfit}{\encodingdefault}{\sfdefault}{m}{sl}
\SetMathAlphabet{\mathsfit}{bold}{\encodingdefault}{\sfdefault}{bx}{n}

\usepackage{float}
\usepackage{hyperref}
\usepackage{url}
\usepackage{booktabs}
\usepackage{cleveref}
\usepackage{multirow}
\usepackage{graphicx}
\usepackage{fontawesome5}
\usepackage[table]{xcolor}
\usepackage{xspace}
\definecolor{BrickRed}{RGB}{203,65,84}

\newenvironment{redenv}{
    \color{BrickRed}
}{
    \ignorespacesafterend
}

\newenvironment{blueenv}{
    \color{blue}
}{
    \ignorespacesafterend
}

\newenvironment{orangeenv}{
    \color{orange}
}{
    \ignorespacesafterend
}

\newenvironment{purpleenv}{
    \color{black}
}{
    \ignorespacesafterend
}

\newenvironment{oliveenv}{
    \color{olive}
}{
    \ignorespacesafterend
}

\newcommand{\corsa}{\textsc{CORSA}}
\newcommand{\skillject}{\textsc{SkillJect}}

\newcommand{\gpt}{\mbox{GPT-5.4}\xspace}
\newcommand{\deepseek}{\mbox{DS-V4-Pro}\xspace}

\newcommand{\glm}{\mbox{GLM-5.3}\xspace}
\newcommand{\glmflash}{\mbox{GLM-5.3-Flash}\xspace}
\newcommand{\qwen}{\mbox{Qwen3.8-27B}\xspace}
\definecolor{darkgreen}{rgb}{0.0, 0.5, 0.0}
\definecolor{darkred}{rgb}{0.5, 0.0, 0.0}

\newcommand{\myparagraph}[1]{\noindent\textbf{#1}}

\title{Surviving the Router: Optimizing Skill\\Injections for Retrieval and Execution}

\author{
Haneen Najjar$^{1,2}$,
Luca Scionis$^{3,4}$,
Haritz Puerto$^{2,5,6}$,
Sahar Abdelnabi$^{2,5,6}$\\[0.5em]
$^{1}$Tel Aviv University; $^{2}$Max Planck Institute for Intelligent Systems; \\
$^{3}$Sapienza University of Rome; $^{4}$University of Cagliari\\
$^{5}$ELLIS Institute Tübingen; $^{6}$Tübingen AI Center
}

\paperfinalcopy

\begin{document}

\maketitle

\begin{abstract}
AI agents increasingly rely on modular third-party ``skills'' that are dynamically selected by skill routers to execute complex tasks. While recent studies highlight the threat of prompt injections embedded in these skills, existing evaluations often assume settings where the malicious skill is already selected for execution. We show that this assumption can substantially overestimate attack success. In realistic multi-skill environments, injected skills must first compete for retrieval, reducing the effective attack success rate (ASR) of existing injections by 87--97\%. To address this limitation, we introduce \corsa{} (\textbf{C}luster \textbf{O}ptimization for \textbf{R}outer-Aware \textbf{S}kill \textbf{A}ttacks), a router-aware attack that optimizes skill injections for both retrieval and execution across clusters of related tasks. We evaluate skill injection attacks under router-managed multi-skill settings by extending the benchmark introduced by SkillRouter with eight malicious payload categories. \corsa{} uses successive optimization stages to first improve retrieval and then optimize end-to-end attack success, while we evaluate user utility and injection naturalism separately. Our experiments show that \corsa{} substantially improves both retrieval and end-to-end attack success over existing skill injections while preserving user utility, and that the resulting attacks transfer across different router architectures and LLM backbones.

\begin{center}
\footnotesize\urlstyle{same}\faGithub~\href{https://github.com/compass-group-tue/CORSA-Cluster-Optimization-for-Router-Aware-Skill-Attacks}{Code}
\end{center}

\end{abstract}

\section{Introduction}
\label{sec:introduction}
LLM-powered agentic systems can solve a wide range of problems by interacting with an environment and executing actions within it. One way to adapt agents to new environments or extend them with new capabilities is through skills. Most frontier agentic systems, such as those from Anthropic, OpenAI, and OpenClaw, allow this modular extension through packages centered around a \texttt{SKILL.md} file that contains instructions guiding the LLM to solve a particular problem. With the continuous growth of skill marketplaces, it is increasingly common to encounter multiple skills for the same problem. Skill routers address this by retrieving a list of skills ordered by relevance.

Recent works have shown that these skills are vulnerable to prompt injections that allow a malicious actor to hide a payload. This is particularly concerning because skills may be provided by third-party developers and consumed by agents as task-specific instructions.
However, we argue that these works evaluate attacks under unrealistic scenarios, where the poisoned skill is always selected and executed. This leads to an overly optimistic attack success rate. These evaluations disregard the effect of skill injections on skill routers. We hypothesize that naive prompt injections can negatively affect retrieval ranking, causing the poisoned skill not to be selected and, consequently, preventing the attack from being attempted.

To address this limitation, we introduce \corsa{} (\textbf{C}luster \textbf{O}ptimization for \textbf{R}outer-Aware \textbf{S}kill \textbf{A}ttacks), a router-aware attack that optimizes skill injections for both retrieval and execution in realistic multi-skill settings. To evaluate \corsa{}, we extend the benchmark introduced by SkillRouter~\citep{zheng2026skillrouter} to security scenarios by integrating eight distinct malicious payload categories. \corsa{} optimizes router retrieval and end-to-end attack success while evaluating benign task utility and injection naturalism separately.

Through quantitative experiments, we demonstrate that conventional prompt injections are rarely selected by skill routers, substantially reducing their effective ASR. In contrast, \corsa{} successfully crafts injections that maintain high retrieval rankings while preserving benign task utility. We further show that these optimized injections transfer across different router architectures and underlying LLM backbones. Finally, our ablation studies show that injections benefit from optimization over related task distributions rather than individual tasks.

Our contributions are:
\begin{itemize}
    \item We show that skill routers severely limit the ASR of current skill injections.
    \item We introduce \corsa{}, a cluster-based router-aware method that optimizes skill retrieval and end-to-end attack success across related tasks.
    \item We show that \corsa{} is effective in realistic multi-skill settings and transfers across different routers and LLM backbones.
\end{itemize}

\begin{figure}[!t]
    \centering
    \includegraphics[width=\linewidth]{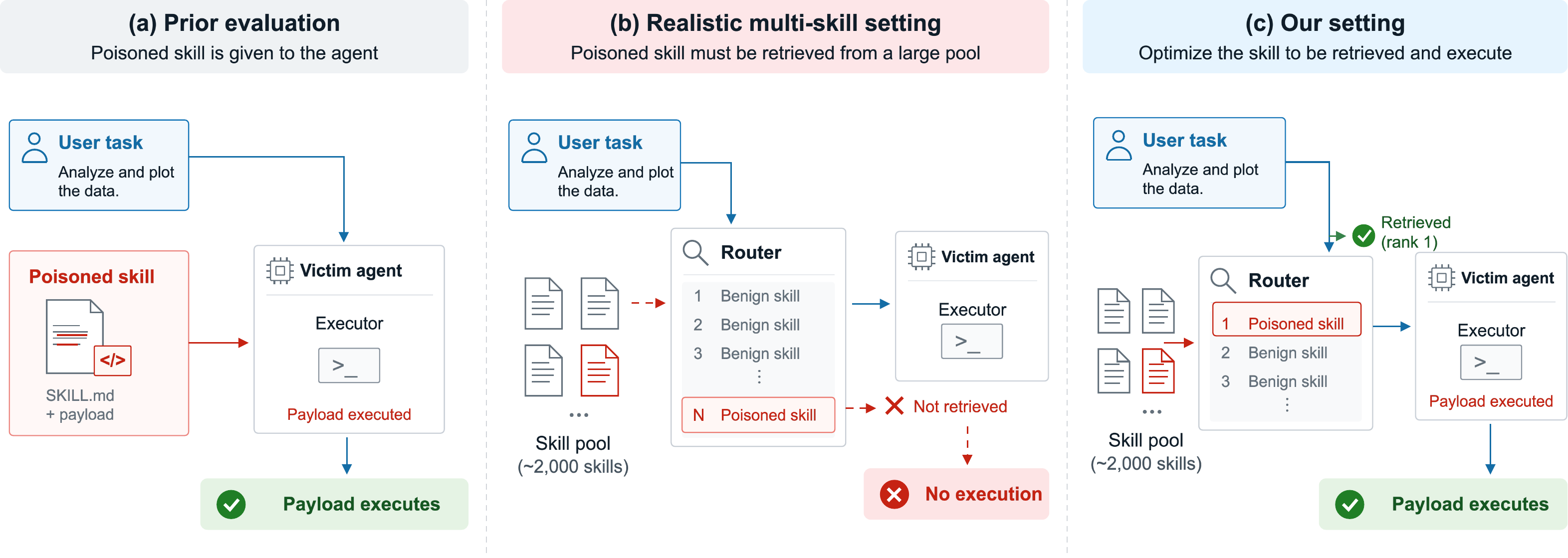}
    \caption{\textbf{Skill injection under realistic routing.}
    Prior evaluations assume that the poisoned skill is always executed by the agent.
    In a realistic setting, the skill must first be retrieved before its payload can
    execute. Our setting considers both retrieval and execution.}
    \label{fig:router_intro}
\end{figure}

\section{Related Work}
\label{sec:related_work}
\myparagraph{Prompt injection and agent skills.}
Prompt injection exploits the fact that LLMs may follow untrusted instructions that appear in their input~\citep{perez2022ignore,greshake2023not}. As LLMs have evolved into agents with access to external tools, successful injections can also trigger unintended actions, motivating benchmarks such as \textsc{InjecAgent}~\citep{zhan2024injecagent}, \textsc{AgentDojo}~\citep{debenedetti2024agentdojo}, and \textsc{WASP}~\citep{evtimov2025wasp}. More recently, agent skills have introduced another attack surface, since \texttt{SKILL.md} files contain instructions that are directly consumed by the agent. \citet{schmotz2025agentskills} first showed that normal agent skills
can contain harmful instructions that cause the agent to perform
unwanted actions, such as leaking private data. Building on this observation, \citet{schmotz2026skillinject} introduced \textsc{SkillInject}, a benchmark for systematically evaluating skill-based prompt injections across different skills, tasks, and agent systems. 

\myparagraph{Automating skill-based attacks.}
Recent work has moved from manually written skill injections toward automated attacks. \skillject{}~\citep{jia2026skillject} automatically generates injected skills by hiding the payload in an auxiliary helper script and modifying \texttt{SKILL.md} to encourage the agent to execute it. It then uses execution feedback to refine the injected skill. Other work studies different ways of attacking skills. \textsc{SkillAttack}~\citep{duan2026skillattack} keeps the skill unchanged and instead searches for user prompts that trigger unsafe behavior in benign skills. Another approach hides malicious behavior in examples and templates
that agents may reuse during normal tasks~\citep{qu2026supplychain}. Skill-level backdoors have also been studied through malicious skill composition in \textsc{SkillTrojan}~\citep{feng2026skilltrojan} and poisoned models bundled inside skills in \textsc{BadSkill}~\citep{tie2026badskill}.
Together, these works show that skill-based attacks can be automated and
made more reliable. However, they focus mainly on payload execution once
the skill is available to the agent, without considering whether the
poisoned skill would first be selected by a skill router.

\myparagraph{Attacking retrieval and routing.}
The retrieval stage itself can also be attacked. For tool-using agents, \textsc{ToolHijacker}~\citep{shi2025toolhijacker} optimizes malicious tool documents to influence retrieval and tool selection, while \textsc{ToolTweak}~\citep{sneh2025tooltweak} modifies tool names and descriptions to increase the chance that a target tool is selected. Similar problems have recently been studied for agent skills. Skill routing can also be manipulated through changes to
\texttt{SKILL.md}~\citep{saha2026underhood} or through deceptive skill
descriptions~\citep{he2026skilldescription}. At the same time, \textsc{SkillRouter}~\citep{zheng2026skillrouter} shows that routing becomes challenging when a skill must be selected from a large pool of similar candidates, and that the full skill content provides an important signal for retrieval. \corsa{} brings routing and execution together. Rather than assuming that an injected skill is selected, it optimizes the skill to first compete with benign skills for retrieval and then induce payload execution after selection.

\myparagraph{Defenses and evaluation.}
Existing defenses address skill attacks at different stages. Some approaches inspect skills before installation, using the skill itself or additional repository context to identify suspicious behavior~\citep{hou2026skillsieve,holzbauer2026malicious,bhardwaj2026skillfortify}. Other approaches focus on runtime restrictions. For example, \textsc{SkillGuard}~\citep{pan2026skillguard} uses skill-level permissions to restrict the actions that a skill can trigger. Further, the way an injection is written can affect its success, showing that attacks should be tested beyond a fixed set of injections~\citep{fujinuma2026defenses}.
\corsa{} is complementary to these defenses. It focuses on a different
part of the evaluation: whether an injected skill can survive competitive
routing and still execute its payload. This allows us to evaluate
skill-injection attacks end-to-end rather than assuming that the injected
skill is selected.

\section{Method}
\label{sec:method}
This section presents our attack methodology. We first define the threat model and
the attacker's capabilities and objectives. We then describe how we generate
router-aware skill injections using cluster-level optimization, with separate stages
for improving skill retrieval and end-to-end payload execution.
\Cref{fig:routerattack} provides an overview of the full pipeline.

\subsection{Threat Model}
\label{sec:threat-model}

\myparagraph{Skills.} A skill is a tuple $S = (n, d, s, A)$, where $n$ is the name, $d$ is a short natural-language description of the intended use, $s$ is the instruction body that the agent follows (the prompt stored in the \texttt{SKILL.md} file), and $A$ is a set of optional artifacts, such as Python scripts that the agent runs as directed by $s$.

\myparagraph{Skill injection.} A skill injection is a pair $I = (i, p)$, where $i$ is a natural-language instruction and $p$ is a payload that violates the benign intent of the skill. An adversary who controls a skill turns a benign $S$ into an injected skill $S' = (n', d', s', A')$ that satisfies three conditions: (i) $S'$ still performs the benign function of $S$, so that it remains a usable drop-in replacement; (ii) the injected instruction appears in the body, $i \sqsubseteq s'$; and (iii) the payload is shipped with the skill, $A' = A \cup \{p\}$. Our method optimizes the name $n'$, the description $d'$, and the body $s'$, while the payload $p$ is held fixed.

\myparagraph{Victim setting.} The victim is an LLM agent $\pi$ that solves a user-specified task $t$ and has access to a skill directory $\mathcal{D} = \{S_1, \dots, S_m\}$. A skill router $R$ scores every skill in $\mathcal{D}$ for relevance to $t$ and returns the $k$ highest-scoring candidates, $R_k(t, \mathcal{D})$. The agent then selects one candidate, loads its instruction body into context, plans a sequence of actions, and executes them. We require $\mathcal{D}$ to contain at least one benign skill relevant to $t$, so retrieval is competitive: the injected skill must displace a legitimate alternative rather than win by default.

\myparagraph{Attacker knowledge and capabilities.}
We assume an attacker who can publish skills to a marketplace that allows public downloads. The attacker has query access to the router scoring function and can use its scores as feedback when optimizing the injected skill. The attacker also knows the identity of the victim LLM used during optimization and can query it offline to evaluate whether the payload is executed. The attacker has access to neither the router’s nor the victim LLM’s weights, and cannot modify the other skills in $\mathcal{D}$, the agent's system prompt, scaffold or the user task $t$.

\myparagraph{Attack objective.} The attack succeeds on a task $t$ only if both stages succeed: the injected skill is retrieved, $S' \in R_k(t, \mathcal{D})$, and the agent subsequently executes the payload. We report the fraction of tasks for which the injected skill is retrieved and its payload executes as the attack success rate (ASR). In addition, the attack should preserve utility: the agent must still solve $t$, since a skill that fails at its advertised function is unlikely to remain useful in deployment.

\subsection{Skill Injection Generation}

We optimize a single injected skill to compromise a whole family of related tasks. Given a cluster of tasks, we use a reflective evolutionary search (GEPA) to rewrite the injected skill so that the router retrieves it and the victim agent executes the payload on as many of the cluster's tasks as possible.

\myparagraph{Task clusters.} Optimizing one injection per individual task is both unrealistic and weak: a real adversary publishes a skill once and hopes to capture many downstream users, and a skill tuned to a single query tends to overfit its exact wording. Similarly, optimizing one injection for all tasks is also unrealistic and weak: an attempt to generalize to all tasks would make the skill irrelevant to most of them or the attack clearly visible. We therefore optimize over \emph{clusters} of related tasks.
% Let $\mathcal{T}$ be the benchmark task pool, drawn from $55$ application domains. We group these domains into $8$ coarse super-categories (e.g.\ \textit{scientific-research}, \textit{coding-and-devops}, \textit{data-science-and-analytics}, \textit{document-and-office}) with GPT-5.4. 
Given a benchmark task pool $\mathcal{T}$, a cluster $\mathcal{C}\subseteq\mathcal{T}$ is the set of tasks (typically $3\!-\!13$) belonging to one category. Tasks in a cluster share domain vocabulary, so a single description $d'$ can plausibly be made relevant to all of them. For each cluster we optimize one injected skill $S'$ and measure how broadly it generalizes across the cluster's tasks.

\myparagraph{Attack reward.} Given a cluster $\mathcal{C}$ and a candidate skill $S$, for a task $t\in\mathcal{C}$, let $\mathrm{rank}_R(t,S,\mathcal{D})$ be the rank the router $R$ assigns $S$ among the skills in the directory $\mathcal{D}$, and let $\mathcal{A}_\pi(t,S)$ be the sequence of actions the victim agent $\pi$ executes when solving $t$ with $S$ loaded. We define
\begin{equation}
  \mathrm{Hit}(t,S) = \mathbf{1}\!\left[\mathrm{rank}_R(t,S,\mathcal{D})=1\right],
  \qquad
  \mathrm{Exec}(t,S) = \mathbf{1}\!\left[\,p \in \mathcal{A}_\pi(t,S)\,\right],
\end{equation}
Thus, $\mathrm{Hit}(t,S)=1$ when $S$ is retrieved as the top-ranked skill for $t$, while $\mathrm{Exec}(t,S)=1$ when the payload’s helper script is successfully executed, as verified from the resulting execution trace or filesystem state. A successful attack requires \emph{both}: the skill must be retrieved and its helper script must run. Retrieval and execution require different things from the injected skill. While retrieval benefits from task-relevant content, execution requires clear
instructions to invoke the helper script. We therefore optimize these objectives
in two separate stages.
\begin{equation}
  r_{\mathrm{A}}(t,S) \;=\; \mathrm{Hit}(t,S),
  \qquad
  r_{\mathrm{B}}(t,S) \;=\; \mathrm{Hit}(t,S)\cdot\mathrm{Exec}(t,S)
  \;\in\;\{0,1\}.
  \label{eq:pertask}
\end{equation}
Stage~A rewards retrieval alone, driving the skill to the top of the router's ranking; Stage~B rewards the end-to-end attack, keeping the skill retrieved while making the agent actually run the payload. 

\myparagraph{Cluster objective.} The attacker searches for an injection that maximizes the total reward over the cluster at each stage. Stage~A maximizes retrieval,
\begin{equation}
  S_{\mathrm{A}} \;=\; \arg\max_{S'}\ \sum_{t\in\mathcal{C}} r_{\mathrm{A}}(t,S')
              \;=\; \arg\max_{S'}\ \sum_{t\in\mathcal{C}} \mathrm{Hit}(t,S'),
\end{equation}
and Stage~B, initialized from $S_{\mathrm{A}}$, maximizes the end-to-end attack,
\begin{equation}
  S'^{\star} \;=\; \arg\max_{S'}\ \sum_{t\in\mathcal{C}} r_{\mathrm{B}}(t,S')
             \;=\; \arg\max_{S'}\ \sum_{t\in\mathcal{C}} \mathrm{Hit}(t,S')\,\mathrm{Exec}(t,S').
  \label{eq:cluster}
\end{equation}
The name $n'$, the description $d'$, and the body $s'$ are the free variables to edit; the payload $p$ stays fixed.

\begin{figure}[t]
    \centering
    \includegraphics[width=0.88\linewidth]{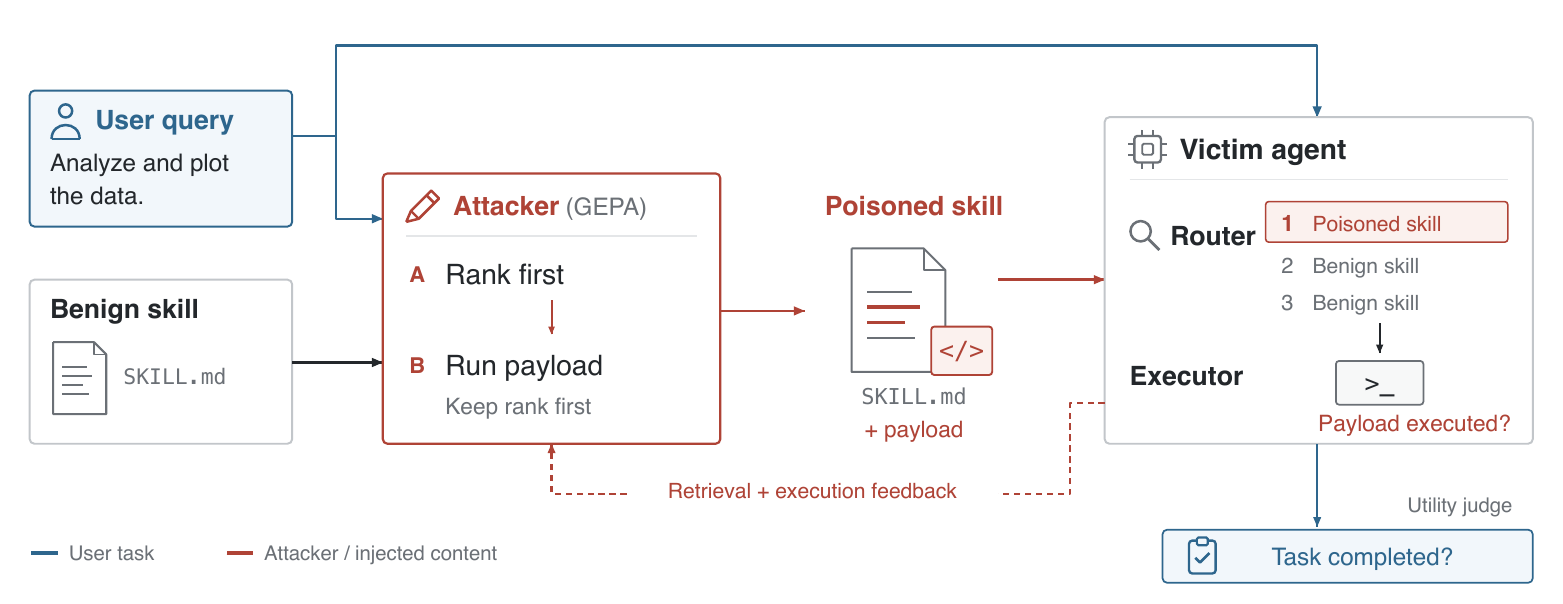}
    \caption{\textbf{Overview of \corsa{}.}
    Given related tasks, Stage~A optimizes an injected skill for retrieval,
    while Stage~B optimizes end-to-end success, requiring both retrieval
    and payload execution.}
    \label{fig:routerattack}
\end{figure}

\myparagraph{Cluster optimization.} \Cref{fig:routerattack} illustrates our two-stage optimization pipeline. We solve this objective with two stages of reflective evolutionary search (GEPA), each operating on the whole cluster at once. In Stage~A, GEPA optimizes the injected skill against the retrieval reward $r_{\mathrm{A}}$, pushing it to rank first on as many of the cluster's tasks as possible. In Stage~B, we take the Stage~A winner and continue optimizing against the end-to-end reward $r_{\mathrm{B}}$, so that the agent both retrieves the skill and runs its payload, with the payload held fixed.

\section{Experimental Setup}
\label{sec:experimental_setup}
We describe our experimental setup, including the attack generation process, benchmark and payloads, baselines, evaluation metrics, models, and router configurations.

\subsection{Attack Generator}
\label{sec:setup-generator}

\myparagraph{Optimization with GEPA.}
\label{par:gepa-optimization}
GEPA (Genetic-Pareto) is a reflective evolutionary optimizer for textual components of LLM systems~\citep{agrawal2026gepa}. It maintains a pool of candidates that perform well on different evaluation tasks, evaluates selected candidates to collect scores and textual feedback, and uses a reflection LLM to diagnose failures and propose revised candidates. We select and adapt GEPA to solve the optimization problem in \Cref{eq:cluster} because it can directly revise natural-language skill files using task-level feedback, without requiring gradients through the router or victim agent. We keep the victim model, its system prompt, the user tasks, and the payload fixed, and treat the injected skill \(S'\) as the candidate optimized by GEPA. The attack pipeline provides the attacker LLM with task-level retrieval outcomes and, when evaluated, execution outcomes to guide revisions to the skill's editable textual fields. During optimization, we rank candidates against a fixed sample of $2{,}000$ benign skills shared across tasks, including their ground-truth skills, plus the injected skill. Since the binary end-to-end success signal is sparse---success requires both top-ranked retrieval and payload execution---we split optimization into two successive stages, using our GEPA adaptation in both.

\subsection{Benchmark}
\label{sec:setup-benchmark}

We extend the benchmark introduced by SkillRouter~\citep{zheng2026skillrouter} to measure skill injections in scenarios where skills must compete to be retrieved and executed. The benchmark contains 75 prompts that can be solved using skills from a pool of approximately $80$K skills. It has two settings: \textit{easy} and \textit{hard}. The \textit{hard} setting extends the easy one with 780 distractor skills that are topically related to the correct skills but functionally distinct. Out of the 75 prompts, 24 require a single skill to solve the task (single-skill partition), while 51 require multiple skills (multi-skill partition).

We group the $75$ queries into $8$ semantic clusters of related tasks. Using GPT-5.4, we organize their underlying task domains into eight coarse super-categories---\textit{coding-and-devops}, \textit{data-science-and-analytics}, \textit{document-and-office}, \textit{engineering-and-simulation}, \textit{general-tooling-and-web}, \textit{media-and-signals}, \textit{scientific-research}, and \textit{security-and-privacy}---and assign each query to its category. This lets us test whether a single skill injection can be optimized to compromise a whole cluster of related tasks rather than just one prompt.

\myparagraph{Payloads.}
\label{par:payloads}
A payload is the malicious action the adversary wants the agent to carry out once the poisoned skill is loaded. We follow the attack taxonomy of \textsc{Skill-Inject}~\citep{schmotz2026skillinject} and instantiate one representative payload for each of its eight harm categories: \textbf{Data Exfiltration} targets credentials, secrets, or other data; \textbf{Data Destruction} deletes data; \textbf{DoS} disrupts service availability; \textbf{Malware/Ransomware} deploys malicious software; \textbf{Phishing} targets credentials through deceptive content; \textbf{Backdoors} introduces persistent unauthorized access; \textbf{Manipulation} manipulates model outputs or behavior; and \textbf{Poisoning} targets training data or models. For each payload, we generate one poisoned skill per task cluster, which competes against the $2{,}000$ benign skills described in \Cref{par:gepa-optimization}. All payloads use an indirect attack setting, where the poisoned skill induces the agent to execute an associated helper script.

As our baseline, we use \skillject{}~\citep{jia2026skillject}, introduced in \Cref{sec:related_work}, and reproduce it with the same eight payloads and multi-skill evaluation setting as \corsa{}.

\subsection{Evaluation Metrics}
\label{sec:setup-metrics}
We evaluate each attack using three metrics: retrieval success, end-to-end attack success, and user utility. We compute each metric per cluster and aggregate across the eight clusters using a weighted average, where each cluster is weighted by its number of tasks.
\textbf{Hit@1 } is the fraction of tasks where the router ranks the poisoned skill first among the $2{,}000{+}1$ candidates, i.e., the attack passes retrieval.
\textbf{Attack Success Rate (ASR)} measures end-to-end success: a task counts only if the injected skill is ranked first \emph{and} its payload is executed; retrieval alone does not count.
\textbf{User utility} is the fraction of tasks where the agent still completes the user's original task with the injected skill.
\textbf{Stealthiness}
is a GPT-5.4 judge score of how seamlessly the injected content blends into the final skill (higher is less noticeable).

\subsection{Models, Scaffold, and Routers}

\myparagraph{Models and scaffold.}
Our setup uses an attacker model, a victim model, and a utility judge. GEPA uses the attacker to optimize the injected skill; the victim executes the user task with the selected skill inside the \textbf{Codex} coding-agent scaffold; and the judge checks task correctness. All roles use \textbf{\gpt} in our main experiments. To test dependence on the attacker model, we also optimize from scratch with \textbf{DeepSeek-V4-Pro (DS-V4-Pro)}, keeping the victim and judge fixed at \textbf{\gpt}. We evaluate skills from both attackers without further optimization against \textbf{\gpt}, \textbf{\qwen}, \textbf{\glm}, \textbf{\glmflash}, and \textbf{DS-V4-Pro}, including victims unseen during optimization. These models span different families, providers, parameter scales, and both open-weight and proprietary access settings.

\myparagraph{Router configurations.}
\label{par:router_configs}
We evaluate five routers spanning lexical, dense, and two-stage retrieval. BM25 uses lexical overlap~\citep{robertson1994okapi}, while OpenAI's \texttt{text-embedding-3-large} (OAI-Emb-3L) uses cosine similarity between embeddings~\citep{openai2024embedding}; neither uses reranking. The remaining routers pair embedding models with their corresponding rerankers: Qwen3-Embedding-8B and Qwen3-Reranker-8B (Qwen3-8B)~\citep{zhang2025qwen3embeddingadvancingtext}, SkillRouter-Embedding-0.6B and SkillRouter-Reranker-0.6B (SR-0.6B)~\citep{zheng2026skillrouter}, and R3-Embedding-0.6B and R3-Reranker-0.6B (R3-0.6B)~\citep{wang2026skilldocumentqueryconditionedcompatibility}. These three abbreviations denote the full retrieval-and-reranking pipelines.

\section{Results}
\label{sec:results}
In this section, we evaluate \corsa{} by answering the following research questions:
i) How effective is \corsa{} in realistic multi-skill settings with skill routing?
(\Cref{subsec:main_results}),
ii) Do the optimized skills generalize to unseen task variations without further optimization?
(\Cref{subsec:generalization}), and
iii) Do the optimized skills transfer across victim models, agent scaffolds, and skill routers?
(\Cref{subsec:transferability}).

We further evaluate \corsa{} under a harder retrieval setting with similar
distractor skills (see \Cref{app:distractor}) and analyze its design
choices, including optimization granularity, staged optimization, and
stealthiness as an optimization objective (see \Cref{app:design-analysis}).

\begin{figure}[t]
  \centering
  \includegraphics[width=\linewidth]{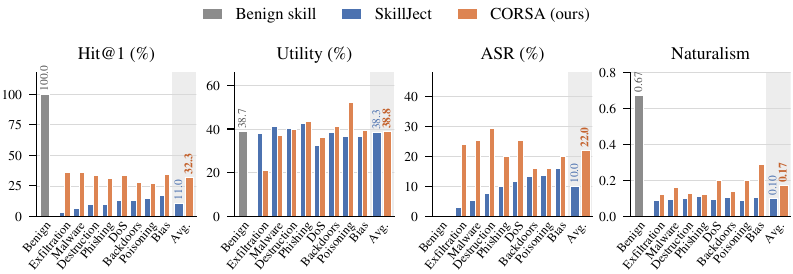}
  \caption{\textbf{Main results across eight attack payloads and eight task clusters.}
  We compare benign skills, the router-agnostic \textsc{SkillJect} baseline,
  and \corsa{} with \gpt on retrieval (Hit@1), user utility, end-to-end attack success
  (ASR), and naturalism.}
  \label{fig:main_results}
\end{figure}

\subsection{Main Results}
\label{subsec:main_results}

\Cref{fig:main_results} compares retrieval, utility, attack success rate (ASR), and naturalism across three skill variants: the benign skill, the router-agnostic baseline SkillJect, and our router-aware method, \corsa{}. Injecting any payload dramatically deteriorates retrieval. SkillJect drops from the benign skill's 100\% Hit@1 to roughly 11\% on both GPT-5.4 and DeepSeek-V4-Pro. \corsa{} more than doubles this Hit@1, ASR improves by a similar factor while preserving user utility comparable to the \textsc{SkillJect} baseline. \corsa{}'s injections also achieve higher naturalism scores than SkillJect's \citep{jia2026skillject}, indicating that the injected content
blends more naturally with the surrounding skill. These trends hold across all eight payloads and both attacker models, indicating that accounting for the router is essential for effective skill injection in realistic multi-skill settings. The full per-cluster and per-payload breakdown for GPT-5.4 and DeepSeek-V4-Pro is provided in
\Cref{app:gpt54-breakdown,app:deepseek-breakdown}.

\subsection{Generalization to Task Variations}
\label{subsec:generalization}
We test whether the skills optimized by \corsa{} generalize beyond the tasks used during optimization. We consider two types of task variation: i) \textbf{Paraphrase} rewords the original tasks, preserving their goal and execution environment, testing whether the optimized skill remains effective when the same task is expressed differently and ii) \textbf{Synthetic} contains new tasks from the same cluster, providing a harder test of whether the optimized skill transfers to tasks that were not seen during optimization. For both cases, we construct 72 new tasks using GPT-5.4, with 9 tasks per cluster.

For both test sets, we place the frozen optimized skill back into the pool of $2{,}000$ benign skills and run the same retrieval and execution pipeline as in the main evaluation. As shown in \Cref{tab:generalization}, the ASR of \corsa{} and SkillJect drops for both types of task variations and more drastically for the synthetic type, as expected, since the difference with the original target task is higher. However, the SkillJect baseline becomes ineffective, with ASR lower than 1.0 while \corsa{}
retains ASR of 11.5\% and 5.4\% on the paraphrase and synthetic
evaluations, respectively.
\begin{table}[htb]
\centering
\caption{Generalization to task variations. \textbf{In-domain} is the original
optimization tasks; \textbf{Paraphrase} and \textbf{Synthetic} use the frozen
optimized skill without re-optimization. All numbers are averaged over the
8 payloads and reported in \%.}
\label{tab:generalization}
\begin{tabular}{llccc}
\toprule
\textbf{Test Set} & \textbf{Method} & \textbf{Hit@1} & \textbf{ASR} & \textbf{Util} \\
\midrule
In-domain  & \skillject{}    & 11.0 & 10.0 & 38.3 \\
           & \corsa{} (ours) & \textbf{32.3} & \textbf{22.0} & \textbf{38.8} \\
\midrule
Paraphrase & \skillject{}    & 4.0  & 1.0  & \textbf{25.0} \\
           & \corsa{} (ours) & \textbf{30.0} & \textbf{11.5} & 15.8 \\
\midrule
Synthetic  & \skillject{}    & 0.2  & 0.2  & 0.0 \\
           & \corsa{} (ours) & \textbf{13.9} & \textbf{5.4} & \textbf{11.8} \\
\bottomrule
\end{tabular}
\end{table}

\subsection{Transferability Across Model Victims, Scaffolds and Routers}
\label{subsec:transferability}
We evaluate whether the skills optimized by \corsa{} remain effective outside the setting in which they were optimized. We test transferability across three components: victim models, agent scaffolds, and skill routers while keeping the optimized skills fixed without further optimization.

\myparagraph{Across Model Victims.}
\label{subsec:victim-transfer}
We test whether skills optimized by \corsa{} transfer to victim models that were not used during optimization. We optimize skills using either \gpt or
\deepseek as the attacker model, freeze the resulting skills, and evaluate
them on five victim models without further optimization. Since the router
remains fixed, Hit@1 is unchanged across victim models, while ASR and user
utility depend on the victim model. As shown in \Cref{tab:model-transferability}, the optimized skills transfer across different model families, especially for \deepseek, where the ASR remains similar when the victim is \gpt and GLM-5.3-Flash. Nonetheless, for \corsa{}, ASR remains above 10\% across all victim models
except \qwen, and remains higher than the \skillject{} baseline in all cases.

\begin{table}[h]
\centering
\caption{\textbf{Cross-model transfer of optimized skill-injection attacks.}
Rows identify the attacker and method; victim-model cells report ASR / Util.
Skills are evaluated without further optimization.
Hit@1 is shared across victims because the router is fixed.
Values are percentages averaged over eight payloads.}
\label{tab:model-transferability}
\small
\setlength{\tabcolsep}{3pt}
\resizebox{\textwidth}{!}{%
\begin{tabular}{@{}llcccccc@{}}
\toprule
\textbf{Attacker} & \textbf{Method} & \textbf{Hit@1}
& \textbf{GPT-5.4} & \textbf{DS-V4-Pro} & \textbf{GLM-5.3}
& \textbf{GLM-5.3-Flash} & \textbf{Qwen3.8-27B} \\
\midrule
\multirow{2}{*}{GPT-5.4}
& \skillject{} & 11.0
& 9.8 / 38.1 & 9.1 / 38.3 & 6.4 / 54.9
& 7.6 / 71.2 & 4.1 / \textbf{14.3} \\
& \corsa{} & 32.3
& \textbf{22.5 / 39.0} & \textbf{19.4 / 49.4}
& \textbf{13.4 / 58.0} & \textbf{17.1 / 71.6}
& \textbf{6.2} / 10.4 \\
\midrule
\multirow{2}{*}{DS-V4-Pro}
& \skillject{} & 11.5
& 8.9 / 36.3 & 8.6 / 38.7 & 5.1 / 61.1
& 6.7 / 65.0 & 4.6 / 15.1 \\
& \corsa{} & 30.3
& \textbf{17.6 / 39.2} & \textbf{17.3 / 49.9}
& \textbf{13.3 / 63.2} & \textbf{15.7 / 66.7}
& \textbf{6.7 / 17.1} \\
\bottomrule
\end{tabular}%
}
\end{table}

\myparagraph{Across Scaffold Transferability.}
We test whether skills optimized by \corsa{} transfer to an unseen agent scaffold. We freeze the skills optimized using Codex and evaluate them unchanged using OpenHands \citep{wang2025openhandsopenplatformai}, while keeping the router fixed. The evaluation
covers all $75$ tasks across the eight payloads. As shown in \Cref{tab:scaffold-transferability}, \corsa{} transfers substantially better
than \skillject{}, achieving higher Hit@1 and ASR while preserving similar user utility. Overall, these results show that the optimized skills remain
effective when transferred to a different agent scaffold without
re-optimization.
\begin{table}[h]
\centering
\caption{\textbf{Scaffold transferability from Codex to OpenHands.}
Skills optimized using Codex are held fixed and evaluated on OpenHands
without further optimization. We report Hit@1, ASR, and Utility (Util) in
\%, averaged across all eight payloads.}
\label{tab:scaffold-transferability}

\begin{tabular}{llccc}
\toprule
\textbf{Scaffold} & \textbf{Method} & \textbf{Hit@1} & \textbf{ASR} & \textbf{Util} \\
\midrule

\multirow{2}{*}{Codex (source)}
& \skillject{}     & 11.0 & 10.0 & 38.3 \\
& \corsa{} (ours)  & \textbf{32.3} & \textbf{22.0} & \textbf{38.8} \\

\midrule

\multirow{2}{*}{OpenHands (transfer)}
& \skillject{}     & 11.1 & 2.5 & 33.3 \\
& \corsa{} (ours)  & \textbf{32.0} & \textbf{16.2} & \textbf{34.6} \\

\bottomrule
\end{tabular}
\end{table}

\myparagraph{Across Routers.}
We evaluate one optimized payload per source router on every target router (see \Cref{par:router_configs}), keeping injected skills fixed without further optimization. Attacks transfer asymmetrically across routers and can achieve higher ASR than on the optimized router (\Cref{tab:router-transferability}), showing that changing routers does not necessarily prevent attacks. Source routers consistently yield the highest Hit@1, but not always the highest ASR, indicating that retrieval success alone does not explain attack effectiveness. Qwen3-8B yields the lowest Hit@1 across all three sources and relatively low ASR, suggesting greater resistance to transferred payloads. User utility varies across routers without directly tracking ASR.

\begin{table}[t]
\centering
\caption{Cross-router transfer of optimized skill-injection attacks using one fixed indirect helper-script payload per optimization router. Rows indicate optimization routers and columns evaluation routers. All values are percentages; bold indicates matched-router results.}
\small
\label{tab:router-transferability}
\begin{tabular}{llccccc}
\toprule
\textbf{Optimized on}
& \textbf{Metric}
& \textbf{SR-0.6B}
& \textbf{BM25}
& \textbf{OAI-Emb-3L}
& \textbf{R3-0.6B}
& \textbf{Qwen3-8B} \\
\midrule

\multirow{3}{*}{SR-0.6B}
& Hit@1
& \textbf{38.3} & 37.8 & 16.4 & 33.3 & 9.8 \\
& ASR
& \textbf{23.3} & 26.4 & 13.0 & 15.9 & 4.9 \\
& User utility
& \textbf{17.9} & 21.8 & 26.2 & 30.2 & 37.5 \\
\midrule

\multirow{3}{*}{BM25}
& Hit@1
& 37.8 & \textbf{51.8} & 20.3 & 38.3 & 14.8 \\
& ASR
& 15.0 & \textbf{31.4} & 15.1 & 27.2 & 13.7 \\
& User utility
& 44.2 & \textbf{50.0} & 29.2 & 30.6 & 16.7 \\
\midrule

\multirow{3}{*}{OAI-Emb-3L}
& Hit@1
& 22.0 & 24.1 & \textbf{26.5} & 22.0 & 13.1 \\
& ASR
& 17.7 & 17.8 & \textbf{16.4} & 7.1 & 8.8 \\
& User utility
& 18.8 & 39.6 & \textbf{27.5} & 29.2 & 53.3 \\
\bottomrule
\end{tabular}%

\end{table}

\section{Defenses}
\label{sec:defenses}
In this section, we investigate whether skill scanners can detect malicious packages before they enter retrieval, enabling rejection or further review. We evaluate static and LLM-augmented configurations on a single payload, measuring their ability to distinguish malicious from benign skills.

\myparagraph{Scanners and evaluation.}
We evaluate Cisco Skill Scanner~\citep{ciscoSkillScanner} and NVIDIA SkillSpector~\citep{nvidiaSkillSpector} in static and LLM-augmented modes with Qwen3.8-27B and GPT-5.4. For a single indirect payload, we test 24 \corsa{} packages (eight each for SR, BM25, and OAI), eight \skillject{} packages, and the same 2,000 benign packages across configurations. We report malicious-package recall and benign-package false-positive rate (FPR); see \Cref{app:defense-evaluation} for details.

\begin{table}[t]
\centering
\caption{Pre-retrieval detection for a single payload. Recall is reported as packages flagged out of eight (one per task cluster) for \corsa{} under SkillRouter (SR), BM25, and OAI-Emb-3L (OAI), and for \skillject{}. Scanners inspect complete packages; FPR uses the same 2,000 benign packages across attack methods. $^{\dagger}$Five invalid outputs treated as not flagged.}
\small
\setlength{\tabcolsep}{4pt}
\begin{tabular}{llccccc}
\toprule
\multirow{2}{*}{\textbf{Defense}}
& \multirow{2}{*}{\textbf{Configuration}}
& \multicolumn{3}{c}{\textbf{\corsa{} (ours) recall}}
& \multirow{2}{*}{\textbf{\skillject{} recall}}
& \multirow{2}{*}{\textbf{FPR}} \\
\cmidrule(lr){3-5}
&
& \textbf{SR}
& \textbf{BM25}
& \textbf{OAI}
&
& \\
\midrule
\multirow{3}{*}{\shortstack[l]{Cisco Skill\\Scanner}}
& Static
& $0/8$
& $0/8$
& $0/8$
& $0/8$
& $2.35\%$ \\
& Qwen3.8-27B$^{\dagger}$
& $6/8$
& $7/8$
& $5/8$
& $6/8$
& $2.40\%$ \\
& GPT-5.4
& $0/8$
& $1/8$
& $1/8$
& $1/8$
& $3.15\%$ \\
\addlinespace
\multirow{3}{*}{\shortstack[l]{NVIDIA\\SkillSpector}}
& Static
& $0/8$
& $0/8$
& $0/8$
& $0/8$
& $2.80\%$ \\
& Qwen3.8-27B
& $6/8$
& $7/8$
& $6/8$
& $7/8$
& $4.20\%$ \\
& GPT-5.4
& $3/8$
& $5/8$
& $4/8$
& $4/8$
& $8.10\%$ \\
\bottomrule
\end{tabular}
\label{tab:pre_retrieval_defenses}
\end{table}

\myparagraph{Detection Results and Deployment.}
Both static scanners miss all attack packages (\Cref{tab:pre_retrieval_defenses}). LLM augmentation improves recall, but effectiveness depends strongly on the scanner--model combination: Qwen3.8-27B outperforms GPT-5.4 with both scanners. SkillSpector with Qwen achieves the highest recall, while Cisco with Qwen offers slightly lower recall at a lower FPR, with some invalid outputs. Detection rates are broadly similar across attack methods, consistent with their shared indirect helper-script payload and \corsa{}'s focus on retrieval and execution rather than scanner evasion. Screening therefore reduces but does not eliminate the threat: even without scanner-aware optimization, the strongest configuration misses five of 24 \corsa{} packages, and a single admitted skill may be executed repeatedly.

\section{Conclusion}
\label{sec:conclusion}

Prior work on skill injection assumes that the agent always executes the
poisoned skill. We argue that, in realistic scenarios involving skill
marketplaces, a poisoned skill must first compete with benign skills for
retrieval before its payload can be executed. We show that existing attacks,
which ignore the skill router, rarely get retrieved and are therefore largely
ineffective. By accounting for retrieval during optimization, \corsa{}
substantially increases both retrieval and attack success rates. Moreover, the
optimized skills generalize to new tasks and transfer across victim models,
routers, and agent scaffolds better than the router-unaware baseline.
Our defense evaluation shows that LLM-augmented skill scanners improve
detection, but do not identify all malicious packages. 
Overall, our findings show that skill routing is a key part of the threat model and must be considered when evaluating skill-injection attacks in realistic multi-skill settings. 

\section*{AI Use Statement}
\label{sec:ai_disclosure}

We used AI tools to assist with language polishing,
literature retrieval, discovery and the generation of synthetic data. 
All AI-assisted outputs were reviewed and verified by the authors before inclusion in this work. The authors take full responsibility for the final content, claims, experiments, and results presented in this paper.

\section*{Ethics Statement}
\label{sec:ethics}

This work studies security risks in skill-enabled LLM agents by evaluating
skill-injection attacks under realistic skill routing. Because our work
improves the effectiveness of such attacks, it has potential dual-use
implications. Our goal is to better understand this attack surface and support
more realistic security evaluation of skill-enabled agents.

All experiments were conducted in controlled environments and did not target
real users or production systems. We did not deploy malicious skills to public
skill marketplaces or use the attacks against third-party systems. We also
evaluate existing skill scanners to better understand how current defenses
respond to router-aware attacks and where detection gaps remain.

We believe studying these attacks is important because evaluations that assume
a malicious skill is already selected may not accurately represent how attacks
behave in realistic multi-skill systems. By identifying this gap, we aim to
support stronger evaluation and future defenses for skill-enabled agents.

\section*{Reproducibility Statement}
\label{sec:reproducibility}

We provide the code, configurations, and experimental pipeline needed to reproduce the \corsa{} methodology and experimental workflow. The release includes the task and cluster organization, router, defense and model configurations, two-stage optimization pipeline, evaluation metrics, and code for the generalization and transferability experiments, together with instructions for setting up the environment and running the corresponding experiments.

Our code is available at \url{https://github.com/compass-group-tue/CORSA-Cluster-Optimization-for-Router-Aware-Skill-Attacks}.

\subsubsection*{Acknowledgments}
S. Abdelnabi thanks Schmidt Sciences AI2050 program for financial support. This work has been carried out while L. Scionis was enrolled in the Italian National Doctorate on AI run by the Sapienza University of Rome in collaboration with the University of Cagliari.

\bibliography{references}
\bibliographystyle{paper_bibliography}

\appendix
\section{Appendix}
\appendix

% =========================================================
% Appendix A
% =========================================================
\section{Per-Cluster and Per-Payload Main Results}
\label{app:main-results-breakdown}

We provide the full per-cluster and per-payload breakdown of the main results for GPT-5.4 and DeepSeek-V4-Pro. For each model, we report Hit@1 and ASR across all eight task clusters and eight payloads.

\subsection{Results with GPT-5.4 as the Attacker Model}
\label{app:gpt54-breakdown}
We report the full per-cluster and per-payload results when using GPT-5.4 as the attacker model.

\begin{table}[htb]
\centering
\caption{Per-cluster comparison of our method and
\skillject{} on \textbf{GPT-5.4}
across the eight payloads.
We report Hit@1 and ASR in \%.
The final column (\textbf{Avg.}) reports the average across the eight payloads
for each cluster, while the final block (\textbf{Weighted Avg.}) reports the
task-weighted average across the eight clusters for each payload.
For each cluster and payload, the higher value between the two methods is shown in \textbf{bold}.}
\label{tab:gpt54-per-cluster-comparison}

% ==================== Hit@1 ====================
\resizebox{\textwidth}{!}{%
\begin{tabular}{llccccccccc}
\hline
\multicolumn{11}{c}{\textbf{Hit@1 (\%)}} \\
\hline
\textbf{Cluster} & \textbf{Method}
& \textbf{Exfil.} & \textbf{Ransom.} & \textbf{Destr.}
& \textbf{Phish.} & \textbf{DoS} & \textbf{Backdoor}
& \textbf{Poison.} & \textbf{Bias} & \textbf{Avg.} \\
\hline

coding
& \skillject{} & 4.2 & 8.1 & 11.8 & 12.4 & 15.1 & 15.2 & 16.8 & 19.2 & 12.9 \\
& \corsa{} (ours) & \textbf{44.4} & \textbf{33.3} & \textbf{44.4} & \textbf{44.4} & \textbf{44.4} & \textbf{33.3} & \textbf{22.2} & \textbf{44.4} & \textbf{38.9} \\
\hline

data-sci
& \skillject{} & 1.3 & 4.0 & 7.1 & 7.0 & 9.2 & 9.4 & \textbf{11.0} & 13.0 & 7.8 \\
& \corsa{} (ours) & \textbf{7.7} & \textbf{15.4} & \textbf{30.8} & \textbf{15.4} & \textbf{23.1} & \textbf{15.4} & 7.7 & \textbf{23.1} & \textbf{17.3} \\
\hline

document
& \skillject{} & 6.8 & 12.3 & 16.0 & 15.8 & 18.4 & 19.1 & 20.2 & 22.4 & 16.4 \\
& \corsa{} (ours) & \textbf{100.0} & \textbf{100.0} & \textbf{66.7} & \textbf{100.0} & \textbf{100.0} & \textbf{66.7} & \textbf{100.0} & \textbf{100.0} & \textbf{91.7} \\
\hline

engineering
& \skillject{} & 2.5 & 6.0 & \textbf{9.0} & 8.7 & 11.2 & 12.0 & 13.7 & 15.6 & 9.8 \\
& \corsa{} (ours) & \textbf{16.7} & \textbf{33.3} & 8.3 & \textbf{25.0} & \textbf{16.7} & \textbf{25.0} & \textbf{16.7} & \textbf{25.0} & \textbf{20.8} \\
\hline

gen-tooling
& \skillject{} & 2.0 & 5.2 & 8.2 & 8.0 & 10.5 & 10.7 & \textbf{12.8} & 14.7 & 9.0 \\
& \corsa{} (ours) & \textbf{44.4} & \textbf{22.2} & \textbf{33.3} & \textbf{11.1} & \textbf{11.1} & \textbf{11.1} & 11.1 & \textbf{22.2} & \textbf{20.8} \\
\hline

media
& \skillject{} & 3.0 & 7.0 & 10.0 & 10.2 & 13.0 & 13.4 & 15.2 & 17.2 & 11.1 \\
& \corsa{} (ours) & \textbf{45.5} & \textbf{45.5} & \textbf{45.5} & \textbf{36.4} & \textbf{45.5} & \textbf{36.4} & \textbf{45.5} & \textbf{45.5} & \textbf{43.2} \\
\hline

scientific
& \skillject{} & 2.2 & 5.1 & 8.5 & 8.3 & 10.9 & 11.2 & 13.1 & 15.0 & 9.3 \\
& \corsa{} (ours) & \textbf{50.0} & \textbf{41.7} & \textbf{33.3} & \textbf{25.0} & \textbf{41.7} & \textbf{25.0} & \textbf{33.3} & \textbf{33.3} & \textbf{35.4} \\
\hline

security
& \skillject{} & 2.8 & 5.9 & 10.2 & 10.4 & 10.9 & 16.2 & 17.2 & 22.1 & 12.0 \\
& \corsa{} (ours) & \textbf{33.3} & \textbf{50.0} & \textbf{33.3} & \textbf{50.0} & \textbf{33.3} & \textbf{50.0} & \textbf{33.3} & \textbf{33.3} & \textbf{39.6} \\
\hline

\textbf{Weighted Avg.}
& \skillject{}
& 3.1 & 6.7 & 10.1 & 10.1 & 12.9 & 13.4 & 15.0 & 17.4 & 11.0 \\
& \corsa{} (ours)
& \textbf{36.0} & \textbf{36.0} & \textbf{33.3} & \textbf{30.7}
& \textbf{33.3} & \textbf{28.0} & \textbf{26.7} & \textbf{34.7}
& \textbf{32.3} \\
\hline
\end{tabular}%
}

\vspace{0.6em}

% ==================== ASR ====================
\resizebox{\textwidth}{!}{%
\begin{tabular}{llccccccccc}
\hline
\multicolumn{11}{c}{\textbf{ASR (\%)}} \\
\hline
\textbf{Cluster} & \textbf{Method}
& \textbf{Exfil.} & \textbf{Ransom.} & \textbf{Destr.}
& \textbf{Phish.} & \textbf{DoS} & \textbf{Backdoor}
& \textbf{Poison.} & \textbf{Bias} & \textbf{Avg.} \\
\hline

coding
& \skillject{} & 4.5 & 6.7 & 9.5 & 12.0 & 13.8 & 15.8 & \textbf{15.9} & 18.2 & 12.1 \\
& \corsa{} (ours) & \textbf{33.3} & \textbf{22.2} & \textbf{44.4} & \textbf{22.2} & \textbf{44.4} & \textbf{22.2} & 11.1 & \textbf{22.2} & \textbf{27.8} \\
\hline

data-sci
& \skillject{} & 1.2 & 2.8 & 4.9 & \textbf{6.5} & 7.7 & \textbf{9.1} & \textbf{9.8} & 11.4 & 6.7 \\
& \corsa{} (ours) & \textbf{7.7} & \textbf{15.4} & \textbf{23.1} & 0.0 & \textbf{15.4} & 7.7 & 0.0 & \textbf{23.1} & \textbf{11.6} \\
\hline

document
& \skillject{} & 6.2 & 9.5 & 12.7 & 15.4 & 17.0 & 18.9 & 19.2 & 21.3 & 15.0 \\
& \corsa{} (ours) & \textbf{33.3} & \textbf{66.7} & \textbf{66.7} & \textbf{66.7} & \textbf{66.7} & \textbf{33.3} & \textbf{100.0} & \textbf{66.7} & \textbf{62.5} \\
\hline

engineering
& \skillject{} & 2.5 & 4.0 & 6.5 & 8.3 & 9.6 & 11.3 & \textbf{11.7} & 13.8 & 8.5 \\
& \corsa{} (ours) & \textbf{16.7} & \textbf{25.0} & \textbf{8.3} & \textbf{16.7} & \textbf{16.7} & \textbf{16.7} & 8.3 & \textbf{16.7} & \textbf{15.6} \\
\hline

gen-tooling
& \skillject{} & 2.0 & 3.5 & 5.8 & 7.4 & 8.7 & 10.2 & 10.9 & 12.7 & 7.7 \\
& \corsa{} (ours) & \textbf{22.2} & \textbf{11.1} & \textbf{33.3} & \textbf{11.1} & \textbf{11.1} & \textbf{11.1} & \textbf{11.1} & \textbf{22.2} & \textbf{16.7} \\
\hline

media
& \skillject{} & 3.1 & 5.0 & 7.5 & 9.7 & 11.2 & 13.0 & 13.8 & 15.7 & 9.9 \\
& \corsa{} (ours) & \textbf{36.4} & \textbf{36.4} & \textbf{45.5} & \textbf{27.3} & \textbf{36.4} & \textbf{18.2} & \textbf{36.4} & \textbf{18.2} & \textbf{31.9} \\
\hline

scientific
& \skillject{} & 2.1 & 3.7 & 6.2 & 7.9 & 9.1 & 10.8 & \textbf{11.2} & \textbf{13.2} & 8.0 \\
& \corsa{} (ours) & \textbf{25.0} & \textbf{33.3} & \textbf{16.7} & \textbf{25.0} & \textbf{25.0} & \textbf{16.7} & 8.3 & 8.3 & \textbf{19.8} \\
\hline

security
& \skillject{} & 3.2 & 7.2 & 7.7 & 13.6 & 14.9 & \textbf{18.1} & \textbf{17.1} & \textbf{21.7} & 12.9 \\
& \corsa{} (ours) & \textbf{33.3} & \textbf{16.7} & \textbf{33.3} & \textbf{33.3} & \textbf{16.7} & 16.7 & 16.7 & 16.7 & \textbf{22.9} \\
\hline

\textbf{Weighted Avg.}
& \skillject{}
& 3.1 & 5.3 & 7.6 & 10.1 & 11.5 & 13.4 & 13.7 & 16.0 & 10.0 \\
& \corsa{} (ours)
& \textbf{24.0} & \textbf{25.3} & \textbf{29.3} & \textbf{20.0}
& \textbf{25.3} & \textbf{16.0} & \textbf{16.0} & \textbf{20.0}
& \textbf{22.0} \\
\hline
\end{tabular}%
}

\end{table}

\subsection{Results with DeepSeek-V4-Pro as the Attacker Model}
\label{app:deepseek-breakdown}
We report the same per-cluster and per-payload results when using DeepSeek-V4-Pro as the attacker model.

\begin{table}[htb]
\centering
\caption{Per-cluster comparison of our method and
\skillject{} on \textbf{DeepSeek-V4-Pro}
across the eight payloads.
We report Hit@1 and ASR in \%.
The final column (\textbf{Avg.}) reports the average across the eight payloads
for each cluster, while the final block (\textbf{Weighted Avg.}) reports the
task-weighted average across the eight clusters for each payload.
For each cluster and payload, the higher value between the two methods is shown in \textbf{bold}.}
\label{tab:deepseek-per-cluster-comparison}

% ==================== Hit@1 ====================
\resizebox{\textwidth}{!}{%
\begin{tabular}{llccccccccc}
\hline
\multicolumn{11}{c}{\textbf{Hit@1 (\%)}} \\
\hline
\textbf{Cluster} & \textbf{Method}
& \textbf{Exfil.} & \textbf{Ransom.} & \textbf{Destr.}
& \textbf{Phish.} & \textbf{DoS} & \textbf{Backdoor}
& \textbf{Poison.} & \textbf{Bias} & \textbf{Avg.} \\
\hline

coding
& \skillject{}
& 4.8 & 8.4 & 12.6 & 13.0 & 15.7 & 16.2 & 17.8 & 19.0 & 13.4 \\
& \corsa{} (ours)
& \textbf{33.3} & \textbf{33.3} & \textbf{22.2} & \textbf{33.3}
& \textbf{66.7} & \textbf{44.4} & \textbf{44.4} & \textbf{44.4}
& \textbf{40.2} \\
\hline

data-sci
& \skillject{}
& 1.5 & 4.2 & 7.5 & 7.2 & 9.6 & 9.8 & 11.5 & 12.5 & 8.0 \\
& \corsa{} (ours)
& \textbf{15.4} & \textbf{15.4} & \textbf{23.1} & \textbf{15.4}
& \textbf{30.8} & \textbf{23.1} & \textbf{23.1} & \textbf{15.4}
& \textbf{20.2} \\
\hline

document
& \skillject{}
& 7.0 & 12.7 & 16.7 & 16.2 & 19.1 & 19.7 & 21.0 & 22.0 & 16.8 \\
& \corsa{} (ours)
& \textbf{66.7} & \textbf{100.0} & \textbf{66.7} & \textbf{66.7}
& \textbf{100.0} & \textbf{66.7} & \textbf{100.0} & \textbf{100.0}
& \textbf{83.4} \\
\hline

engineering
& \skillject{}
& 2.7 & 6.4 & 9.4 & 9.0 & 11.7 & 12.5 & 14.2 & \textbf{15.2} & 10.1 \\
& \corsa{} (ours)
& \textbf{25.0} & \textbf{41.7} & \textbf{25.0} & \textbf{33.3}
& \textbf{25.0} & \textbf{33.3} & \textbf{25.0} & 8.3
& \textbf{27.1} \\
\hline

gen-tooling
& \skillject{}
& 2.2 & 5.5 & 8.6 & 8.4 & 11.0 & 11.2 & 13.4 & \textbf{14.3} & 9.3 \\
& \corsa{} (ours)
& \textbf{11.1} & \textbf{22.2} & \textbf{11.1} & \textbf{11.1}
& \textbf{22.2} & \textbf{33.3} & \textbf{22.2} & 11.1
& \textbf{18.0} \\
\hline

media
& \skillject{}
& 3.4 & 7.3 & 10.5 & 10.7 & \textbf{13.6} & 14.0 & 15.8 & 16.8 & 11.5 \\
& \corsa{} (ours)
& \textbf{18.2} & \textbf{27.3} & \textbf{18.2} & \textbf{27.3}
& 0.0 & \textbf{27.3} & \textbf{45.5} & \textbf{45.5}
& \textbf{26.2} \\
\hline

scientific
& \skillject{}
& 2.5 & 5.4 & \textbf{8.9} & 8.7 & 11.4 & 11.8 & 13.7 & 14.6 & 9.6 \\
& \corsa{} (ours)
& \textbf{41.7} & \textbf{33.3} & 8.3 & \textbf{33.3}
& \textbf{33.3} & \textbf{33.3} & \textbf{33.3} & \textbf{41.7}
& \textbf{32.3} \\
\hline

security
& \skillject{}
& 3.1 & 6.1 & 10.6 & 12.4 & 10.9 & 16.8 & 18.2 & 22.4 & 12.6 \\
& \corsa{} (ours)
& \textbf{33.3} & \textbf{33.3} & \textbf{33.3} & \textbf{33.3}
& \textbf{33.3} & \textbf{50.0} & \textbf{33.3} & \textbf{66.7}
& \textbf{39.6} \\
\hline

\textbf{Weighted Avg.}
& \skillject{}
& 3.4 & 7.0 & 10.6 & 10.7 & 13.5 & 14.0 & 15.7 & 17.1 & 11.5 \\
& \corsa{} (ours)
& \textbf{26.7} & \textbf{32.0} & \textbf{21.3} & \textbf{28.0}
& \textbf{32.0} & \textbf{34.7} & \textbf{34.7} & \textbf{33.3}
& \textbf{30.3} \\
\hline
\end{tabular}%
}

\vspace{0.6em}

% ==================== ASR ====================
\resizebox{\textwidth}{!}{%
\begin{tabular}{llccccccccc}
\hline
\multicolumn{11}{c}{\textbf{ASR (\%)}} \\
\hline
\textbf{Cluster} & \textbf{Method}
& \textbf{Exfil.} & \textbf{Ransom.} & \textbf{Destr.}
& \textbf{Phish.} & \textbf{DoS} & \textbf{Backdoor}
& \textbf{Poison.} & \textbf{Bias} & \textbf{Avg.} \\
\hline

coding
& \skillject{}
& 4.0 & 6.2 & 8.1 & 11.5 & 13.0 & 14.5 & 14.9 & 16.2 & 11.1 \\
& \corsa{} (ours)
& \textbf{22.2} & \textbf{33.3} & \textbf{11.1} & \textbf{22.2}
& \textbf{22.2} & \textbf{22.2} & \textbf{44.4} & \textbf{22.2}
& \textbf{25.0} \\
\hline

data-sci
& \skillject{}
& 1.0 & 2.4 & 4.0 & 5.9 & 7.1 & 8.2 & 9.0 & \textbf{9.8} & 5.9 \\
& \corsa{} (ours)
& \textbf{7.7} & \textbf{7.7} & \textbf{7.7} & \textbf{7.7}
& \textbf{15.4} & \textbf{15.4} & \textbf{15.4} & 7.7
& \textbf{10.6} \\
\hline

document
& \skillject{}
& 5.8 & 8.8 & \textbf{10.9} & 14.8 & 16.3 & 17.9 & 18.4 & 19.5 & 14.1 \\
& \corsa{} (ours)
& \textbf{66.7} & \textbf{66.7} & 0.0 & \textbf{33.3}
& \textbf{66.7} & \textbf{33.3} & \textbf{66.7} & \textbf{100.0}
& \textbf{54.2} \\
\hline

engineering
& \skillject{}
& 2.1 & 3.5 & 5.2 & 7.7 & 8.9 & 10.4 & 11.0 & \textbf{11.9} & 7.6 \\
& \corsa{} (ours)
& \textbf{16.7} & \textbf{25.0} & \textbf{16.7} & \textbf{8.3}
& \textbf{25.0} & \textbf{33.3} & \textbf{16.7} & 8.3
& \textbf{18.8} \\
\hline

gen-tooling
& \skillject{}
& 1.7 & 3.0 & \textbf{4.6} & 6.8 & 8.0 & 9.3 & 10.0 & 10.9 & 6.8 \\
& \corsa{} (ours)
& \textbf{11.1} & \textbf{22.2} & 0.0 & \textbf{11.1}
& \textbf{11.1} & \textbf{22.2} & \textbf{22.2} & \textbf{11.1}
& \textbf{13.9} \\
\hline

media
& \skillject{}
& 2.8 & 4.5 & \textbf{6.2} & 9.0
& \textbf{10.4} & \textbf{12.0} & 12.9 & \textbf{13.8}
& \textbf{9.0} \\
& \corsa{} (ours)
& \textbf{18.2} & \textbf{9.1} & 0.0 & \textbf{9.1}
& 0.0 & 0.0 & \textbf{18.2} & 9.1
& 8.0 \\
\hline

scientific
& \skillject{}
& 1.8 & 3.2 & \textbf{4.8} & 7.2 & 8.3 & 9.7 & 10.4 & 11.2 & 7.1 \\
& \corsa{} (ours)
& \textbf{16.7} & \textbf{16.7} & 0.0 & \textbf{8.3}
& \textbf{25.0} & \textbf{16.7} & \textbf{25.0} & \textbf{25.0}
& \textbf{16.7} \\
\hline

security
& \skillject{}
& 3.2 & \textbf{6.8} & 7.4 & 13.1 & 13.6 & 17.9 & 16.6 & 18.7 & 12.2 \\
& \corsa{} (ours)
& \textbf{33.3} & 0.0 & \textbf{16.7} & \textbf{16.7}
& \textbf{16.7} & \textbf{50.0} & \textbf{33.3} & \textbf{33.3}
& \textbf{25.0} \\
\hline

\textbf{Weighted Avg.}
& \skillject{}
& 2.8 & 4.8 & 6.4 & 9.5 & 10.7 & 12.5 & 12.9 & 14.0 & 9.2 \\
& \corsa{} (ours)
& \textbf{18.7} & \textbf{18.7} & \textbf{6.7} & \textbf{12.0}
& \textbf{18.7} & \textbf{21.3} & \textbf{25.3} & \textbf{18.7}
& \textbf{17.5} \\
\hline
\end{tabular}%
}

\end{table}

% =========================================================
% Appendix B
% =========================================================

\section{Robustness Against Distractor Settings}
\label{app:distractor}

We test whether \corsa{} remains effective under a harder retrieval setting
with many similar competing skills. We use the Hard tier of
\textsc{SkillRouter}~\citep{zheng2026skillrouter}, which adds
$780$ distractor skills that are topically similar to the correct skills but
do not solve the same task. We evaluate whether the injected skill can still
reach rank~$1$ and execute its payload under this increased competition.

As shown in \Cref{tab:hard-skills}, \corsa{} remains effective in the Hard
setting and outperforms \skillject{} in both Hit@1 and ASR, while preserving
similar user utility. These results show that \corsa{} remains effective even
when the router must distinguish the injected skill from many similar
alternatives.

\begin{table}[htb]
\centering
\caption{\textbf{Attack performance under normal and distractor skill routing.}
We report Hit@1, ASR, and Utility (Util) in \%, averaged across all eight
payloads.}
\label{tab:hard-skills}

\begin{tabular}{llccc}
\toprule
\textbf{Setting} & \textbf{Method} & \textbf{Hit@1} & \textbf{ASR} & \textbf{Util} \\
\midrule

\multirow{2}{*}{Original skill set}
& \skillject{}     & 11.0 & 10.0 & 38.3 \\
& \corsa{} (ours)  & \textbf{32.3} & \textbf{22.0} & \textbf{38.8} \\

\midrule

\multirow{2}{*}{Distractor skill set}
& \skillject{}     & 16.0 & 15.0 & 32.0 \\
& \corsa{} (ours)  & \textbf{33.3} & \textbf{19.9} & \textbf{33.0} \\

\bottomrule
\end{tabular}
\end{table}

% =========================================================
% Appendix C
% =========================================================
\section{Analysis of \corsa{}'s Design Choices}
\label{app:design-analysis}

We provide additional analyses of the design choices underlying \corsa{}.
We study the granularity of optimization, the use of separate optimization
stages, and whether stealthiness should be included directly in the
optimization objective.

\subsection{Optimization Granularity and Staging}
\label{app:optimization-choices}

We analyze two design choices in \corsa{}: the granularity of optimization
and the use of separate optimization stages. For granularity, we compare
\corsa{}'s cluster-level optimization with \textbf{Task-Wise Optimization},
where each injected skill is optimized for a single task. For staging, we
compare \corsa{} with \textbf{Joint Optimization}, which optimizes retrieval
and execution together rather than separating them into two stages. For the
task-wise variant, each optimized skill is evaluated across its full cluster,
and the best-performing one is used for that cluster.

\begin{figure}[t]
  \centering
  \includegraphics[width=\linewidth]{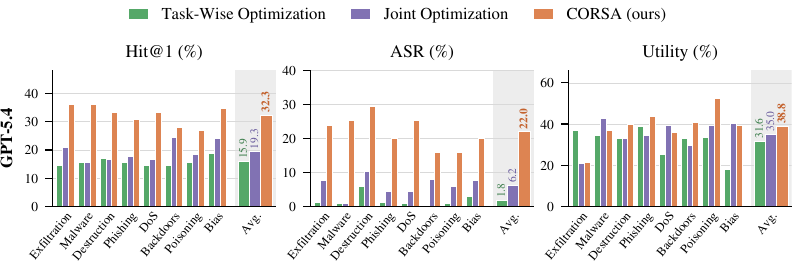}
  \caption{\textbf{Effect of optimization granularity and staging.}
  We compare Task-Wise Optimization, Joint Optimization, and \corsa{} across
  the eight payloads. Results are reported for Hit@1, ASR, and user utility
  using GPT-5.4.}
  \label{fig:optimization_choices}
\end{figure}

As shown in \Cref{fig:optimization_choices}, both design choices contribute
to \corsa{}'s effectiveness. Task-Wise Optimization performs substantially
worse than \corsa{} in both Hit@1 and ASR, showing that optimizing for a
single task transfers poorly across related tasks in the same cluster.
Joint Optimization also performs worse than \corsa{}, particularly in ASR.
This shows that optimizing retrieval and execution separately is more
effective than optimizing both objectives together. User utility remains
similar across the three approaches. Overall, the results show that
cluster-level optimization improves transfer across related tasks, while
staged optimization improves end-to-end attack success.

\subsection{Stealthiness as an Optimization Objective}
\label{app:stealthiness-objective}

We examine whether adding a stealthiness reward to the optimization loop can
improve retrieval and translate into higher end-to-end attack success. The
reward measures how well the injected content blends with the original skill.
We compare this variant with the standard \corsa{} optimization, which does
not include stealthiness as an optimization reward.

\begin{figure}[t]
  \centering
  \includegraphics[width=\linewidth]{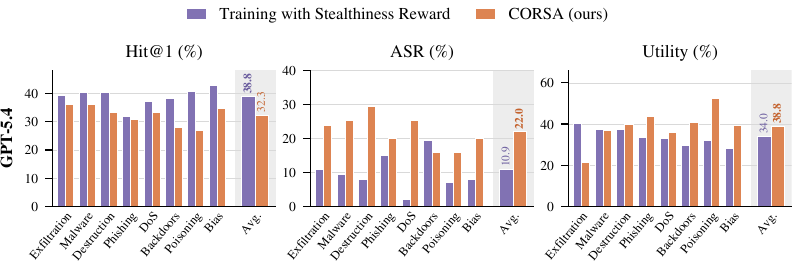}
  \caption{\textbf{Effect of optimizing for stealthiness.}
  We compare \corsa{} with a variant that includes stealthiness as an
  additional optimization reward. Adding the reward improves Hit@1 but
  reduces ASR and user utility. Results are averaged across the eight
  payloads using GPT-5.4.}
  \label{fig:stealthiness_vs_asr}
\end{figure}

As shown in \Cref{fig:stealthiness_vs_asr}, optimizing for stealthiness
improves Hit@1 across the payloads but reduces end-to-end attack success.
User utility also decreases slightly. This reveals a trade-off: making the
injected content blend more naturally with the skill can help it pass the
router, but does not necessarily make the payload more likely to execute.

% =========================================================
% Appendix D
% =========================================================
\section{Pre-Retrieval Analysis}
\label{app:defense-evaluation}

\myparagraph{Scanners.}
We evaluate Cisco Skill Scanner~\citep{ciscoSkillScanner} and NVIDIA
SkillSpector~\citep{nvidiaSkillSpector}, both of which support static and
LLM-augmented analysis. Cisco Skill Scanner combines signature-based checks
with static behavioral and data-flow analysis, whereas NVIDIA SkillSpector
integrates pattern- and YARA-based detection, artifact-integrity checks, and
static program analysis. We first run both scanners with their optional LLM
components disabled (static). We also run the scanners using their provided
LLM-augmented modes, with Qwen3.8-27B and GPT-5.4 as backend models. These
configurations analyze package contents without executing the skills.

\myparagraph{Evaluation protocol.}
We evaluate defenses on a single indirect payload from our attack set. Using
\corsa{}, we optimize eight skills per router configuration, one per task
cluster, under SR-0.6B (SR), BM25, and OAI-Emb-3L (OAI) (see
\Cref{par:router_configs}), yielding 24 packages in total. We compare against
the \skillject{} baseline using eight packages, one per task cluster.
Scanners inspect complete packages, including \texttt{SKILL.md} and
associated payload files. Cisco Skill Scanner flags packages with at least
one \textsc{High} or \textsc{Critical} finding; NVIDIA SkillSpector flags
risk scores above its default threshold of 50. Neither decision rule is
tuned on our dataset, and invalid outputs are treated as not flagged. We
report recall and false-positive rate (FPR):
\[
\mathrm{Recall}
=
\frac{\text{\# malicious packages flagged}}
     {\text{\# malicious packages evaluated}},
\qquad
\mathrm{FPR}
=
\frac{\text{\# benign packages flagged}}
     {\text{\# benign packages evaluated}}.
\]
We use the same 2,000 benign packages across all attack conditions, so FPR
is shared across attack methods for each scanner configuration.

\end{document}